\documentclass[journal]{IEEEtran}
\usepackage[T1]{fontenc}
\usepackage{ifpdf}
\usepackage{cite}
\usepackage[pdftex]{graphicx}
\usepackage{amsmath}
\usepackage{amssymb}
\usepackage{bm}
\usepackage{array}
\usepackage{fixltx2e}
\usepackage{stfloats}

\usepackage{balance} 

\usepackage{mathtools} 

\usepackage{diagbox}

\usepackage{makecell}

\usepackage{threeparttable}

\usepackage{color,soul}
\soulregister\cite7

\usepackage[hidelinks]{hyperref}

\begin{document}
\title{\Large Minimizing the Installation Cost of Ground Stations in Satellite Networks: \\ Complexity, Dynamic Programming and Approximation Algorithm}

\author{Christos~N.~Efrem and~Athanasios~D.~Panagopoulos,~\IEEEmembership{Senior~Member,~IEEE}
\thanks{C. N. Efrem and A. D. Panagopoulos are with the School of Electrical and Computer Engineering, National Technical University of Athens, 15780 Athens, Greece (e-mails: chefr@central.ntua.gr, thpanag@ece.ntua.gr).

This article has been accepted for publication in \textit{IEEE Wireless Communications Letters}, DOI: 10.1109/LWC.2020.3031717. Copyright \textcopyright \ 2020 IEEE. Personal use is permitted, but republication/redistribution requires IEEE permission. See \url{http://www.ieee.org/publications_standards/publications/rights/index.html} for more information.
}}

\markboth{}%
{}

\maketitle

\begin{abstract}
In this letter, we study the optimum selection of ground stations (GSs) in RF/optical satellite networks (SatNets) in order to minimize the overall installation cost under an outage probability requirement, assuming independent weather conditions between sites. First, we show that the optimization problem can be formulated as a binary-linear-programming problem, and then we give a formal proof of its NP-hardness. Furthermore, we design a dynamic-programming algorithm of pseudo-polynomial complexity with global optimization guarantee as well as an efficient (polynomial-time) approximation algorithm with provable performance guarantee on the distance of the achieved objective value from the global optimum. Finally, the performance of the proposed algorithms is verified through numerical simulations.

\end{abstract}


\IEEEpeerreviewmaketitle

\section{Introduction}
\IEEEPARstart{T}{he} availability of satellite networks (SatNets) is heavily affected by atmospheric impairments, especially rain in radio-frequency (RF) and clouds in optical SatNets. Site diversity techniques are able to improve the network availability by mitigating the extremely high attenuation induced by rain and clouds \cite{Panagopoulos}. An optimization method for selecting optical GSs is proposed in \cite{Fuchs}, taking into consideration the single-site availabilities and the spatial-correlation between sites as well. In \cite{Poulenard}, a joint optimization algorithm for the design of optical SatNets is presented, which is divided into two parts: the GS positioning and the backbone network optimization considering the optical fiber cost.

Moreover, \cite{Lyras_GEO} and \cite{Lyras_MEO} present low-complexity heuristic algorithms, which exploit the spatial correlation and the monthly variability of cloud coverage, in order to select the minimum number of GSs in optical SatNets with a geostationary (GEO) or a medium-earth-orbit (MEO) satellite, respectively. A multi-objective optimization approach that achieves various tradeoffs between availability, latency and cost is examined in \cite{Portillo}, so as to determine the optimal location of optical GSs for low-earth-orbit (LEO) SatNets. In addition, as concerns the smart gateway diversity optimization in extremely-high-frequency (EHF) SatNets, \cite{Rossi} presents another multi-objective approach using genetic algorithms. 

Recently, \cite{Gong} provides an efficient gradient-projection method to select a given number of GSs maximizing the availability of free-space optical (FSO) SatNets. Finally, a branch-and-bound (B\&B) algorithm with global optimization guarantee and low average-case complexity is developed in \cite{Efrem} to select the minimum number of GSs under availability requirements for each time period.

In this letter, we develop useful optimization algorithms for selecting GSs with the minimum installation cost satisfying an outage probability constraint. More specifically, the main contributions of this letter are summarized as follows: 
\begin{itemize}
\item Mathematical formulation of the optimization problem in binary-linear-programming form with a rigorous proof of its computational complexity (\textit{NP-hardness}).
\item Design of a \textit{dynamic-programming algorithm with pseudo-polynomial complexity}, which is theoretically guaranteed to find the global optimum. 
\item Design of a \textit{polynomial-time approximation algorithm with provable performance guarantee} on the distance between the objective value of the achieved solution and the global optimum (thus achieving a reasonable performance-complexity tradeoff). 
\item Unlike existing approaches that minimize just the number of GSs (cardinality minimization problem, assuming implicitly the same cost for each GS), the proposed algorithms minimize the overall installation cost allowing \textit{possibly different costs of GSs}. 
\end{itemize}

The remainder of this letter is organized as follows. Section II presents the formulation of the optimization problem with a theoretical proof of its NP-hardness. Subsequently, a global optimization algorithm using dynamic programming is given in Section III, while a polynomial-time approximation algorithm is presented in Section IV. Finally, Section V provides some numerical results and Section VI concludes this letter.

\textit{Mathematical notation}: The set of positive integers is \linebreak denoted by ${\mathbb{Z}_ + } = \{ 1,2,3, \ldots \} $, while ${{\mathbf{0}}_K}$ and ${{\mathbf{1}}_K}$ are respectively the $K$-dimensional all-zeros and all-ones vectors. Moreover, $\left\lfloor  \cdot  \right\rfloor $ and $\left\lceil  \cdot  \right\rceil $ stand for the floor and ceiling functions, respectively.  

\section{Problem Formulation \& NP-Hardness}
Consider an RF/optical SatNet with site diversity, consisting of a GEO satellite and a network of geographically distributed GSs. In particular, $\mathcal{K} = \{ 1,2, \ldots ,K\}$ denotes the set of candidate locations/sites for installing a GS ($K \in {\mathbb{Z}_ + }$). In addition, we assume that: 1) the \textit{network outage probability} is defined as the probability of having \textit{all GSs in outage} and 2) \textit{the distance between any two distinct locations is large enough} so that the spatial correlation between sites can be ignored, without significant error on the calculation of network outage probability; this implies (approximately) \textit{independent weather conditions} between the candidate locations. 

In this context, we study the minimization of the total installation cost of GSs satisfying a given outage probability requirement: 
\begin{subequations} \label{initial_problem}
\begin{alignat}{3}
  & \mathop {\min }\limits_\mathcal{S} & \quad & \sum\limits_{s \in \mathcal{S}} {{c_s}}  \\
  & \text{subject to} & & {P_{{\text{out}}}}(\mathcal{S}) \leq P_{{\text{out}}}^{{\text{th}}} \\
  & & & \mathcal{S} \subseteq \mathcal{K}    
\end{alignat}
\end{subequations}
where $\mathcal{S}$ is the set of selected locations, ${c_k} \in {\mathbb{Z}_ + }$ denotes the cost of installing a GS at the ${k^{{\text{th}}}}$ location , $\forall k \in \mathcal{K}$ (without loss of generality, we assume that ${c_1} \leq {c_2} \leq  \cdots  \leq {c_K}$; this requires an extra complexity of $O(K\log K)$ for sorting the sites in ascending-cost order),\footnote{Note that the coefficient $c_k$ may include the cost of fiber-optic cables needed to connect the $k^{\text{th}}$ GW to the existing access points (points of presence) of the terrestrial backbone network.} ${P_{{\text{out}}}}(\mathcal{S}) = \prod_{s \in \mathcal{S}} {{p_s}}$ is the network outage probability achieved by the set $\mathcal{S}$, with ${p_k} \in (0,1]$ being the outage probability of a GS installed at the ${k^{{\text{th}}}}$ location, $\forall k \in \mathcal{K}$,\footnote{The outage probability of each GS can be obtained from experimental data (when available) or using time-series synthesizers. Moreover, in RF SatNets a GS is in outage when the rain attenuation is higher than a specific threshold \cite{Rossi}, whereas in optical SatNets when experiencing cloud blockage \cite{Fuchs}.} and $P_{{\text{out}}}^{{\text{th}}} \in (0,1]$ is the (network) outage probability threshold. Herein, ${[{p_k}]_{k \in \mathcal{K}}}$ and $P_{{\text{out}}}^{{\text{th}}}$ are defined on an annual basis, and therefore the proposed approach does not take into account the monthly/seasonal variability of weather conditions. Note that in the special case where \linebreak ${c_k} = 1$, $\forall k \in \mathcal{K}$, we have a cardinality minimization problem. 

Afterwards, by introducing the vector ${\mathbf{z}} = [{z_1},{z_2}, \ldots ,{z_K}]$ of binary (0/1) variables (${z_k} = 1$ if and only if $k \in \mathcal{S}$), we can equivalently formulate problem \eqref{initial_problem} as follows (note that $\sum_{s \in \mathcal{S}} {{c_s}}  = \sum_{k \in \mathcal{K}} {{c_k}{z_k}}$ and ${P_{{\text{out}}}}(\mathcal{S}) = \prod_{k \in \mathcal{K}} {{{({p_k})}^{{z_k}}}}$): 
\begin{subequations} \label{equivalent_problem_A}
\begin{alignat}{3}
  & \mathop {\min }\limits_{\mathbf{z}} & \quad & f({\mathbf{z}}) = \sum\limits_{k \in \mathcal{K}} {{c_k}{z_k}}  \\
  & \text{subject to} & & \prod\limits_{k \in \mathcal{K}} {{{({p_k})}^{{z_k}}}}  \leq P_{{\text{out}}}^{{\text{th}}} \\
  & & & {z_k} \in \{ 0,1\}, \; \forall k \in \mathcal{K}    
\end{alignat}
\end{subequations}

Exploiting the fact that $x \leq y$ $\Leftrightarrow$ $\log (x) \leq \log (y)$, $\forall x,y > 0$, the constraint $\prod_{k \in \mathcal{K}} {{{({p_k})}^{{z_k}}}}  \leq P_{{\text{out}}}^{{\text{th}}}$ is equivalent to $\sum_{k \in \mathcal{K}} {{z_k}\log ({p_k})}  \leq \log (P_{{\text{out}}}^{{\text{th}}})$. Consequently, problem \eqref{equivalent_problem_A} can be written as a \textit{binary-linear-programming problem}:
\begin{subequations} \label{equivalent_problem_B}
\begin{alignat}{3}
  & \mathop {\min }\limits_{\mathbf{z}} & \quad & f({\mathbf{z}}) = \sum\limits_{k \in \mathcal{K}} {{c_k}{z_k}}  \\
  & \text{subject to} & & \sum\limits_{k \in \mathcal{K}} {{a_k}{z_k}}  \geq b \\
  & & & {z_k} \in \{ 0,1\}, \; \forall k \in \mathcal{K}    
\end{alignat}
\end{subequations}
where ${a_k} =  - \log ({p_k}) \geq 0$, $\forall k \in \mathcal{K}$, and $b =  - \log (P_{{\text{out}}}^{{\text{th}}}) \geq 0$. Let $\mathcal{F} = \left\{ {{\mathbf{z}} \in {{\{ 0,1\} }^K}: \, \prod_{k \in \mathcal{K}} {{{({p_k})}^{{z_k}}}}  \leq P_{{\text{out}}}^{{\text{th}}}} \right\}$, or equivalently $\mathcal{F} = \left\{ {{\mathbf{z}} \in {{\{ 0,1\} }^K}: \, \sum_{k \in \mathcal{K}} {{a_k}{z_k}}  \geq b} \right\}$, be the feasible set and ${{\mathbf{z}}^ * } \in \mathop {\arg \min }_{\mathbf{z}} \left\{ {f({\mathbf{z}}): \, {\mathbf{z}} \in \mathcal{F}} \right\}$ be a (globally) optimal solution of problem \eqref{equivalent_problem_A}/\eqref{equivalent_problem_B}. Since ${a_k} \geq 0$, $\forall k \in \mathcal{K}$, the following \textit{necessary and sufficient feasibility condition} applies: problem \eqref{equivalent_problem_A}/\eqref{equivalent_problem_B} is feasible (i.e., $\mathcal{F} \neq \emptyset$) \textit{if and only if} $\prod_{k \in \mathcal{K}} {{p_k}} \leq P_{{\text{out}}}^{{\text{th}}}$ or, equivalently, $\sum_{k \in \mathcal{K}} {{a_k}} \geq b$ (i.e., ${{\mathbf{1}}_K} \in \mathcal{F}$).

\vspace{2mm}
\newtheorem{theorem}{Theorem}
\begin{theorem}[NP-hardness] 
The binary-linear-programming problem \eqref{equivalent_problem_B} is NP-hard.
\end{theorem}
\vspace{2mm}

\renewcommand{\IEEEQED}{\IEEEQEDopen}
\begin{IEEEproof}
In order to prove the NP-hardness of problem \eqref{equivalent_problem_B}, it is sufficient to show that a special case of this problem is NP-hard. Firstly, let consider the \textit{0-1 knapsack problem} which is a well-known NP-hard problem \cite{Papadimitriou}:
\begin{subequations} \label{knapsack_problem}
\begin{alignat}{3}
  & \mathop {\max }\limits_{\mathbf{x}} & \quad & \sum\limits_{k \in \mathcal{K}} {{v_k}{x_k}}  \\
  & \text{subject to} & & \sum\limits_{k \in \mathcal{K}} {{w_k}{x_k}}  \leq W \\
  & & & {x_k} \in \{ 0,1\}, \; \forall k \in \mathcal{K}    
\end{alignat}
\end{subequations}
where $W \in {\mathbb{Z}_ + }$ is the knapsack capacity, and ${v_k},{w_k} \in {\mathbb{Z}_ + }$ are the value and weight of the $k^\text{th}$ item, respectively, $\forall k \in \mathcal{K}$. Moreover, applying the \textit{polynomial-time, $\Theta (K)$, variable transformation} ${x_k} = 1 - {z_k}$, $\forall k \in \mathcal{K}$, we get the following equivalent problem:\footnote{Note that the optimum objective values of problems \eqref{knapsack_problem} and \eqref{transformed_knapsack_problem} differ only by a constant, i.e., $\sum_{k \in \mathcal{K}} {{v_k}x_k^ * }  = \sum_{k \in \mathcal{K}} {{v_k}}  - \sum_{k \in \mathcal{K}} {{v_k}z_k^ * }$.} 
\begin{subequations} \label{transformed_knapsack_problem}
\begin{alignat}{3}
  & \mathop {\min }\limits_{\mathbf{z}} & \quad & \sum\limits_{k \in \mathcal{K}} {{v_k}{z_k}}  \\
  & \text{subject to} & & \sum\limits_{k \in \mathcal{K}} {{w_k}{z_k}}  \geq W' \\
  & & & {z_k} \in \{ 0,1\}, \; \forall k \in \mathcal{K}    
\end{alignat}
\end{subequations}
where $W' = \sum_{k \in \mathcal{K}} {{w_k}} - W$.  Without loss of generality, we can assume that the integer $W' \geq 0$; otherwise the optimal solution of problem \eqref{transformed_knapsack_problem} is trivially equal to ${{\mathbf{0}}_K}$. Obviously, the NP-hard problem \eqref{transformed_knapsack_problem} is a subcase of problem \eqref{equivalent_problem_B}, and this completes the proof.
\end{IEEEproof}
\vspace{2mm}

\section{Global Optimization Using Dynamic Programming}
Due to the fact that problem \eqref{equivalent_problem_B} is NP-hard, it cannot be (globally) solved in polynomial time unless P$=$NP. Nevertheless, we can use a powerful optimization technique, namely, \textit{dynamic programming (DP)}, in order to achieve the global minimum with pseudo-polynomial complexity.

DP performs an intelligent enumeration of all the feasible solutions, thus providing a \textit{global optimization guarantee}. In particular, DP follows a \textit{bottom-up approach} by decomposing the problem into ``smaller'' subproblems and combining their optimal solutions (using a \textit{recursive formula}) in order to find an optimal solution to the original problem; this is known as \textit{the principle of optimality} and such problems are said to have \textit{optimal substructure} \cite{Papadimitriou}. Furthermore, DP is a \textit{tabular method} where each subproblem is solved only once and then its solution is stored in a table, so that it can be readily used (without re-computation) by ``larger'' problems when needed.

Let $C$ be an \textit{integer upper bound} on the optimum value of problem \eqref{equivalent_problem_B}, i.e., $f({{\mathbf{z}}^ * }) \leq C$, where $C \in \{ 0,1, \ldots ,\overline C \} $ with $\overline C  = \sum_{k \in \mathcal{K}} {{c_k}} $ (this is \textit{the ``worst'' upper bound} that can be used). In addition, we define the following bivariate function $\forall i \in {\mathcal{K}_0} = \{ 0,1, \ldots ,K\} $ and $\forall j \in {\mathcal{C}_0} = \{ 0,1, \ldots ,C\} $:
\begin{equation} \label{R_function}
\scalebox{0.98}{$
R(i,j) = \mathop {\max }\limits_{{{\mathbf{z}}_\mathcal{I}}} \left\{ {\sum\limits_{k \in \mathcal{I}} {{a_k}{z_k}} : \, \sum\limits_{k \in \mathcal{I}} {{c_k}{z_k}}  = j,\ {{\mathbf{z}}_\mathcal{I}} \in {{\{ 0,1\} }^i}} \right\}
$}
\end{equation}
where $\mathcal{I} = \{ 1,2, \ldots ,i\} $ and ${{\mathbf{z}}_\mathcal{I}} = [{z_1},{z_2}, \ldots ,{z_i}]$, with $i = 0$ $ \Rightarrow $ $\mathcal{I} = \emptyset $ and $\sum_{k \in \emptyset} {{a_k}{z_k}}  = \sum_{k \in \emptyset} {{c_k}{z_k}}  = 0$. If this maximization problem is infeasible, then $R(i,j) =  - \infty $. 

\vspace{2mm}
\begin{theorem}[Computation of the global optimum] 
Assuming that problem \eqref{equivalent_problem_A}/\eqref{equivalent_problem_B} is feasible, its global minimum can be found as follows: $f({{\mathbf{z}}^ * }) = \min \left\{ {j \in {\mathcal{C}_0}: \, R(K,j) \geq b} \right\}$. 
\end{theorem}
\vspace{2mm}

\begin{IEEEproof}
Firstly, observe that when $i = K$, we have $\mathcal{I} = \mathcal{K}$ and ${{\mathbf{z}}_\mathcal{I}} = {\mathbf{z}}$. Secondly, we know that $f({{\mathbf{z}}^ * }) \in {\mathcal{C}_0}$ and $R(K,f({{\mathbf{z}}^ * })) \geq \sum_{k \in \mathcal{K}} {{a_k}z_k^ * }  \geq b$. Now, suppose that $f({{\mathbf{z}}^ * }) \ne {j^ * }$, where ${j^ * } = \min \left\{ {j \in {\mathcal{C}_0}: \, R(K,j) \geq b} \right\}$. Let examine two cases: 1) $f({{\mathbf{z}}^ * }) < {j^ * }$ and 2) $f({{\mathbf{z}}^ * }) > {j^ * }$. In the former case, we would have that $R(K,f({{\mathbf{z}}^ * })) < b$, which leads to a contradiction. Moreover, the latter case contradicts the global optimality of $f({{\mathbf{z}}^ * })$. Hence, $f({{\mathbf{z}}^ * }) = {j^ * }$ and Theorem 2 has been proven. 
\end{IEEEproof}
\vspace{2mm}

Subsequently, we partition the feasible set of problem \eqref{R_function}, by setting ${z_i} = 0$ and ${z_i} = 1$, respectively (note that $\mathcal{I}\backslash \{ i\}  = \{ 1,2, \ldots ,i - 1\} $): 
\begin{equation} \label{problem_with_z0}
\scalebox{0.84}{$
\begin{gathered}
  \mathop {\max }\limits_{{{\mathbf{z}}_\mathcal{I}}} \left\{ {\sum\limits_{k \in \mathcal{I}} {{a_k}{z_k}} : \, \sum\limits_{k \in \mathcal{I}} {{c_k}{z_k}}  = j,\ {{\mathbf{z}}_\mathcal{I}} \in {{\{ 0,1\} }^i},\ {z_i} = 0} \right\} =  \hfill \\
   = \mathop {\max }\limits_{{{\mathbf{z}}_{\mathcal{I}\backslash \{ i\} }}} \left\{ {\sum\limits_{k \in \mathcal{I}\backslash \{ i\} } {{a_k}{z_k}} : \, \sum\limits_{k \in \mathcal{I}\backslash \{ i\} } {{c_k}{z_k}}  = j,\ {{\mathbf{z}}_{\mathcal{I}\backslash \{ i\} }} \in {{\{ 0,1\} }^{i - 1}}} \right\} =  \hfill \\
   = R(i - 1,j) \hfill \\ 
\end{gathered} 
$}
\end{equation}
\begin{equation} \label{problem_with_z1}
\scalebox{0.78}{$
\begin{gathered}
  \mathop {\max }\limits_{{{\mathbf{z}}_\mathcal{I}}} \left\{ {\sum\limits_{k \in \mathcal{I}} {{a_k}{z_k}} : \, \sum\limits_{k \in \mathcal{I}} {{c_k}{z_k}}  = j,\ {{\mathbf{z}}_\mathcal{I}} \in {{\{ 0,1\} }^i},\ {z_i} = 1} \right\} =  \hfill \\
   = {a_i} + \mathop {\max }\limits_{{{\mathbf{z}}_{\mathcal{I}\backslash \{ i\} }}} \left\{ {\sum\limits_{k \in \mathcal{I}\backslash \{ i\} } {{a_k}{z_k}} : \, \sum\limits_{k \in \mathcal{I}\backslash \{ i\} } {{c_k}{z_k}}  = j - {c_i},\ {{\mathbf{z}}_{\mathcal{I}\backslash \{ i\} }} \in {{\{ 0,1\} }^{i - 1}}} \right\} =  \hfill \\
   = {a_i} + R(i - 1,j - {c_i}) \hfill \\ 
\end{gathered} 
$}
\end{equation}

Therefore, we have the following \textit{recursive formula} \linebreak $\forall i \in \mathcal{K} = \{ 1,2, \ldots ,K\} $ and $\forall j \in {\mathcal{C}_0} = \{ 0,1, \ldots ,C\} $:
\begin{equation} \label{recursive_formula}
\scalebox{0.97}{$
R(i,j) = \left\{ \begin{gathered}
  \max \left\{ {R(i - 1,j),{a_i} + R(i - 1,j - {c_i})} \right\},\; {\text{if}} \; j \geq {c_i} \hfill \\
  R(i - 1,j),\; {\text{otherwise}} \hfill \\ 
\end{gathered}  \right.
$}
\end{equation}
with \textit{initial conditions}: a) $R(0,0) = 0$ and b) $R(0,j) =  - \infty $, $\forall j \in \mathcal{C} = \{ 1,2, \ldots ,C\} $. Observe that if $j < {c_i}$, then problem \eqref{problem_with_z1} is definitely infeasible, so $R(i - 1,j - {c_i})= -\infty $; this explains the 2\textsuperscript{nd} branch in \eqref{recursive_formula}. 

Algorithm 1 presents a DP procedure based on the previous analysis. First, we compute the coefficients $[a_k]_{k \in \mathcal{K}}$ and $b$ (line 1), and then a \textit{greedy method} is used in order to calculate the upper bound $C$ (lines 2-5). In essence, this method sequentially selects GSs in ascending-cost order and finds a feasible solution to problem \eqref{equivalent_problem_B}, which is certainly an upper bound on the optimum value. Afterwards, the algorithm stores the $R(i,j)$ values in a $(K + 1) \times (C + 1)$ table, whose entries are computed in row order from left to right (lines 6-15). Moreover, the global optimum can be found by checking the last row, since $f({{\mathbf{z}}^ * }) = {j^ * } = \min \left\{ {j \in {\mathcal{C}_0}: \, R(K,j) \geq b} \right\}$ according to Theorem 2. Finally, an optimal solution can be deduced from the generated table by starting at $R(K,{j^ * })$ and tracing where the optimal values come from (lines 16-23). In particular, if $R(i,j) = R(i - 1,j)$, then $z_i^ *  = 0$, and we continue tracing with $R(i - 1,j)$. Otherwise $z_i^ *  = 1$, and we continue tracing with $R(i - 1,j - {c_i})$. This process is repeated for each $i$ from $K$ down to $1$ (with step $-1$). Therefore, \textit{Algorithm 1 is theoretically guaranteed to find a (globally) optimal solution}.

\begin{table}[!t]
\centering
\renewcommand{\arraystretch}{1.35}
\begin{tabular*}{\columnwidth}{@{}l@{}}
\hline
\textbf{\normalsize{Algorithm 1}} \normalsize{Dynamic Programming (DP)}
\\ \hline
\textbf{Input:} $K \in {\mathbb{Z}_ + }$, ${\mathbf{c}} = [{c_1},{c_2}, \ldots ,{c_K}] \in \mathbb{Z}_ + ^K$ where ${c_1} \leq {c_2} \leq  \cdots  \leq {c_K}$, \\ 
${\mathbf{p}} = [{p_1},{p_2}, \ldots ,{p_K}] \in {(0,1]^K}$, $P_{{\text{out}}}^{{\text{th}}} \in (0,1]$ with $\prod\limits_{k \in \mathcal{K}} {{p_k}}  \leq P_{{\text{out}}}^{{\text{th}}}$ \\
\textbf{Output:} ${{\mathbf{z}}^ * } \in \mathop {\arg \min }\limits_{\mathbf{z}} \left\{ {\sum\limits_{k \in \mathcal{K}} {{c_k}{z_k}} : \, {\mathbf{z}} \in \mathcal{F}} \right\}$ \\
\begin{tabular}{@{}r@{~}l@{}}
1: & ${a_k} \coloneqq  - \log ({p_k})$, $\forall k \in \mathcal{K}$, $b \coloneqq  - \log (P_{{\text{out}}}^{{\text{th}}})$ \\
2: & $A \coloneqq 0$, $C \coloneqq 0$, $k \coloneqq 1$ \\
3: & \textbf{while} $A < b$ \textbf{do}   \hfill $\triangleright$ \textit{Calculation of the upper bound} $C$ \\
4: & ~~$A \coloneqq A + {a_k}$, $C \coloneqq C + {c_k}$, $k \coloneqq k + 1$   \\
5: & \textbf{end while} \\
6: & $R(0,0) \coloneqq 0$, $R(0,j) \coloneqq  - \infty$, $\forall j \in \mathcal{C}$  \\
7: & \textbf{for} $i \coloneqq 1$ \textbf{to} $K$ \textbf{step} $+1$ \textbf{do}  \hfill $\triangleright$ \textit{Computation of the table} $R$ \\
8: & ~~\textbf{for} $j \coloneqq 0$ \textbf{to} $C$ \textbf{step} $+1$ \textbf{do} \\
9: & ~~~~\textbf{if} $j \geq {c_i}$ \textbf{then} \\
10: & ~~~~~~$R(i,j) \coloneqq \max \left\{ {R(i - 1,j),{a_i} + R(i - 1,j - {c_i})} \right\}$ \\ 
11: & ~~~~\textbf{else}  \\
12: & ~~~~~~$R(i,j) \coloneqq R(i - 1,j)$ \\
13: & ~~~~\textbf{end if} \\
14: & ~~\textbf{end for} \\
15: & \textbf{end for}  \\ 
16: & ${j^ * } \coloneqq \min \left\{ {j \in {\mathcal{C}_0}: \, R(K,j) \geq b} \right\}$, $q \coloneqq R(K,{j^ * })$, $j \coloneqq {j^ * }$ \\ 
17: & \textbf{for} $i \coloneqq K$ \textbf{to} $1$ \textbf{step} $-1$ \textbf{do} \;\; \hfill $\triangleright$ \textit{Reconstruction of an optimal solution} \\ 
18: & ~~\textbf{if} $q = R(i - 1,j )$ \textbf{then}  \\
19: & ~~~~$z_i^ * \coloneqq 0$, $q \coloneqq R(i - 1,j )$ \\
20: & ~~\textbf{else} \\
21: & ~~~~$z_i^ * \coloneqq 1$, $q \coloneqq R(i - 1,j - {c_i})$, $j \coloneqq j  - {c_i}$ \\
22: & ~~\textbf{end if} \\
23: & \textbf{end for}  \\
\end{tabular}
\\ \hline
\end{tabular*}
\end{table}

\textit{Complexity of Algorithm 1}: The complexity of computing the coefficients $[a_k]_{k \in \mathcal{K}}$ and $b$ is $\Theta (K)$. Moreover, the greedy method used to find an upper bound on the optimum value requires at most $K$ iterations (since ${{\mathbf{1}}_K} \in \mathcal{F}$), thus having $O(K)$ complexity. In addition, the computation of the table $R$ requires $\Theta (KC)$ arithmetic operations in total. Finally, the computation of $j^*$ requires $O(C)$ comparisons, while the complexity of reconstructing/tracing the solution is $\Theta (K)$ since it starts in row $K$ of the table and moves up one row at each step. Ultimately, \textit{the overall complexity of Algorithm 1 is} $\Theta (KC) = O(K\overline C ) = O({K^2}{c_{\max }})$, because $C \leq \overline C  \leq K{c_{\max }}$ where ${c_{\max }} = \mathop {\max }_{k \in \mathcal{K}} \{ {c_k}\}  = {c_K}$. As a result, the proposed DP algorithm has \textit{pseudo-polynomial time complexity} \cite{Papadimitriou}. 

\textit{Remark 1}: Strictly speaking, Algorithm 1 is an \textit{exponential-time algorithm}, since the size of the input is upper bounded by $O(K\log {c_{\max }}) = O(K\log \overline C )$, because ${c_{\max }} \leq \overline C $. Nevertheless, under certain conditions, this algorithm is practical despite its exponential worst-case complexity. For example, if $\overline C  = O({K^d})$ for some constant $d \geq 0$ (which is usually the case in practice), then the running time of Algorithm 1 will be \textit{polynomial in $K$}. In any case, the optimization problem under consideration does not need to be solved in real time, but during the initial network design.

\textit{Remark 2}: In Algorithm 1, due to the fact that \linebreak $C = \sum_{u \in \mathcal{U}} {{c_u}}$ for some $\mathcal{U} \subseteq \mathcal{K}$ depending on $[a_k]_{k \in \mathcal{K}}$ and $b$, we can divide all coefficients $[c_k]_{k \in \mathcal{K}}$ with their \textit{greatest common divisor} (i.e., ${c'_k} = {c_k}/\zeta \in {\mathbb{Z}_ + } $, $\forall k \in \mathcal{K}$, where $\zeta  = \gcd ({c_1},{c_2}, \ldots ,{c_K}) \in {\mathbb{Z}_ + }$) without altering the set of optimal solutions. In this way, \textit{the complexity of Algorithm 1 can be reduced}, since $C' = \sum_{u \in \mathcal{U}} {{c'_u}}  = C/\zeta \leq C$.

\section{Polynomial-Time Approximation Algorithm}
Subsequently, a practical and efficient (polynomial-time) approximation algorithm with \textit{provable performance guarantee} is given. The design of the approximation algorithm is based on the idea of trading accuracy for running time, thus achieving a \textit{reasonable tradeoff between performance and complexity}. 

The approximation algorithm utilizes Algorithm 1 and is shown in Algorithm 2. Specifically, Algorithm 2 is similar to the fully polynomial-time approximation scheme (FPTAS) for the knapsack problem provided in \cite{Vazirani}, which is inspired by the work of Ibarra and Kim \cite{Ibarra_Kim}. Moreover, note that $\vartheta  > 0$, and therefore ${\tilde c_k} \in {\mathbb{Z}_ + }$, $\forall k \in \mathcal{K}$.

\begin{table}[!t]
\centering
\renewcommand{\arraystretch}{1.35}
\begin{tabular*}{\columnwidth}{@{}l@{}}
\hline
\textbf{\normalsize{Algorithm 2}} \normalsize{DP-based Approximation Algorithm (DPAA)}
\\ \hline
\textbf{Input:} $K \in {\mathbb{Z}_ + }$, ${\mathbf{c}} = [{c_1},{c_2}, \ldots ,{c_K}] \in \mathbb{Z}_ + ^K$ where ${c_1} \leq {c_2} \leq  \cdots  \leq {c_K}$, \\ 
${\mathbf{p}} = [{p_1},{p_2}, \ldots ,{p_K}] \in {(0,1]^K}$, $P_{{\text{out}}}^{{\text{th}}} \in (0,1]$ with $\prod\limits_{k \in \mathcal{K}} {{p_k}}  \leq P_{{\text{out}}}^{{\text{th}}}$, $\epsilon > 0$ \\
\textbf{Output:} ${{\mathbf{\tilde z}}^ * } \in \mathcal{F}$ such that $f({{\mathbf{z}}^ * }) \leq f({{\mathbf{\tilde z}}^ * }) \leq f({{\mathbf{z}}^ * }) + \min (\left\lfloor {\epsilon {c_{\max }}} \right\rfloor ,\overline C )$ \\
\begin{tabular}{@{}r@{~}l@{}}
1: & $\vartheta \coloneqq {\epsilon c_{\max }}/K$, where ${c_{\max }} = \mathop {\max }\limits_{k \in \mathcal{K}} \{ {c_k}\}  = {c_K}$ \\
2: & ${\tilde c_k} \coloneqq \left\lceil {{c_k}/\vartheta } \right\rceil $, $\forall k \in \mathcal{K}$   \\
3: & Run Algorithm 1 with input $[K$, ${\mathbf{\tilde c}}$, ${\mathbf{p}}$, $P_{{\text{out}}}^{{\text{th}}}]$ and return the optimal \\
& solution ${{\mathbf{\tilde z}}^ * }$, where ${\mathbf{\tilde c}} = [{\tilde c_1},{\tilde c_2}, \ldots ,{\tilde c_K}] \in \mathbb{Z}_ + ^K$ $({\tilde c_1} \leq {\tilde c_2} \leq  \cdots  \leq {\tilde c_K})$ \\                                                                                                                                                 
\end{tabular}
\\ \hline
\end{tabular*}
\end{table}

\vspace{2mm}
\begin{theorem}[Performance guarantee] 
Assuming that problem \eqref{equivalent_problem_A}/\eqref{equivalent_problem_B} is feasible, Algorithm 2 takes a parameter $\epsilon > 0$ as input and produces an approximate solution ${{\mathbf{\tilde z}}^ * } \in \mathcal{F}$ such that $f({{\mathbf{z}}^ * }) \leq f({{\mathbf{\tilde z}}^ * }) \leq f({{\mathbf{z}}^ * }) + \min (\left\lfloor {\epsilon {c_{\max }}} \right\rfloor ,\overline C )$, where the term $\min (\left\lfloor {\epsilon {c_{\max }}} \right\rfloor ,\overline C )$ is \textit{the absolute-error bound}. In addition, for any $0 < \epsilon < 1/{c_{\max }}$, Algorithm 2 always finds an optimal solution, i.e., it becomes an exact optimization algorithm.
\end{theorem}
\vspace{2mm}

\begin{IEEEproof}
Obviously, ${{\mathbf{\tilde z}}^ * } \in \mathcal{F}$ and thus $f({{\mathbf{z}}^ * }) \leq f({{\mathbf{\tilde z}}^ * })$. Now, it is sufficient to prove that $f({{\mathbf{\tilde z}}^ * }) \leq f({{\mathbf{z}}^ * }) + \min (\left\lfloor {\epsilon {c_{\max }}} \right\rfloor ,\overline C )$. First, we will show that $f({{\mathbf{\tilde z}}^ * }) \leq f({{\mathbf{z}}^ * }) + \epsilon {c_{\max }}$. Due to the fact that $x \leq \left\lceil x \right\rceil  < x + 1$, we have ${c_k}/\vartheta  \leq {\tilde c_k} < {c_k}/\vartheta  + 1$ $ \Rightarrow $ ${c_k} \leq \vartheta {\tilde c_k} < {c_k} + \vartheta $. Also, let us define the function $g({\mathbf{z}}) = \sum_{k \in \mathcal{K}} {{{\tilde c}_k}{z_k}}$. From $\vartheta {\tilde c_k} < {c_k} + \vartheta $, we deduce that $\vartheta {\tilde c_k}z_k^ *  \leq {c_k}z_k^ *  + \vartheta z_k^ * $, $\forall k \in \mathcal{K}$ (because $z_k^ *  \geq 0$). By taking the sum for all $k \in \mathcal{K}$, we obtain $\vartheta g({{\mathbf{z}}^ * }) \leq f({{\mathbf{z}}^ * }) + \vartheta \sum_{k \in \mathcal{K}} {z_k^ * }  \leq f({{\mathbf{z}}^ * }) + \vartheta K = f({{\mathbf{z}}^ * }) + \epsilon {c_{\max }}$. Since ${{\mathbf{z}}^ * } \in \mathcal{F}$, we conclude that $g({{\mathbf{\tilde z}}^ * }) \leq g({{\mathbf{z}}^ * })$ $ \Rightarrow $ $\vartheta g({{\mathbf{\tilde z}}^ * }) \leq \vartheta g({{\mathbf{z}}^ * })$ because $\vartheta > 0$, and therefore $\vartheta g({{\mathbf{\tilde z}}^ * }) \leq f({{\mathbf{z}}^ * }) + \epsilon {c_{\max }}$. In addition, from ${c_k} \leq \vartheta {\tilde c_k}$ $ \Rightarrow $ ${c_k}\tilde z_k^ *  \leq \vartheta {\tilde c_k}\tilde z_k^ * $, $\forall k \in \mathcal{K}$ (because $\tilde z_k^ *  \geq 0$). By taking the sum for all $k \in \mathcal{K}$ once more, we get $f({{\mathbf{\tilde z}}^ * }) \leq \vartheta g({{\mathbf{\tilde z}}^ * })$. Consequently, $f({{\mathbf{\tilde z}}^ * }) \leq f({{\mathbf{z}}^ * }) + \epsilon {c_{\max }}$. Afterwards, due to the fact that $f({{\mathbf{\tilde z}}^ * })$ and $f({{\mathbf{z}}^ * })$ are \textit{integers}, we have $f({{\mathbf{\tilde z}}^ * }) - f({{\mathbf{z}}^ * }) \leq \left\lfloor {\epsilon {c_{\max }}} \right\rfloor $. Moreover, since $f({{\mathbf{\tilde z}}^ * }) \leq \overline C $ and $f({{\mathbf{z}}^ * }) \geq 0$, we obtain $f({{\mathbf{\tilde z}}^ * }) - f({{\mathbf{z}}^ * }) \leq \overline C $. Hence, $f({{\mathbf{\tilde z}}^ * }) - f({{\mathbf{z}}^ * }) \leq \min (\left\lfloor {\epsilon {c_{\max }}} \right\rfloor ,\overline C )$, because it holds that: $x \leq u$ and $x \leq v$ $ \Leftrightarrow $ $x \leq \min (u,v)$. 

Furthermore, if $0 < \epsilon < 1/{c_{\max }}$ $\Rightarrow$ $0 < \epsilon {c_{\max }} < 1$ $\Rightarrow$ $\left\lfloor {\epsilon {c_{\max }}} \right\rfloor  = 0$, and thus $f({{\mathbf{z}}^ * }) \leq f({{\mathbf{\tilde z}}^ * }) \leq f({{\mathbf{z}}^ * })$ $ \Rightarrow $ $f({{\mathbf{\tilde z}}^ * }) = f({{\mathbf{z}}^ * })$. In other words, for any $0 < \epsilon < 1/{c_{\max }}$, the approximation algorithm will be forced to produce an optimal solution.
\end{IEEEproof}
\vspace{2mm}

\textit{Complexity of Algorithm 2}: The complexity of DPAA is mainly due to Algorithm 1, so it is $O({K^2}{\tilde c_{\max }}) = O({K^2}\left\lceil {K/\epsilon} \right\rceil ) = O({K^3}/\epsilon)$, where ${\tilde c_{\max }} = \mathop {\max }_{k \in \mathcal{K}} \{ {\tilde c_k}\}  = \left\lceil {{c_{\max }}/\vartheta } \right\rceil  = \left\lceil {K/\epsilon} \right\rceil$. As a result, \textit{Algorithm 2 has polynomial complexity in $K$ and $1/\epsilon$}. Observe that, for any fixed $\epsilon>0$, DPAA has cubic complexity $O({K^3})$.

Finally, the performance and complexity of all optimization algorithms are summarized in Table I. The exhaustive search algorithm simply checks all subsets of $\mathcal{K}$ and selects that with the minimum objective value satisfying the outage probability constraint. Therefore, it requires $\sum_{i = 0}^K {{\binom{K}{i}} i} = K{2^{K-1}} = \Theta ({2^K}K)$ arithmetic operations to find the global minimum.

\begin{table}[!t]
\caption{Performance \& Complexity of Optimization Algorithms}
\centering
\renewcommand{\arraystretch}{2.0}
\begin{tabular}{|c|c|c|}
\hline
\makecell{\textbf{Optimization} \\ \textbf{Algorithm}} & \makecell{\textbf{Performance} \\ \textbf{Guarantee}} & \makecell{\textbf{Computational} \\ \textbf{Complexity}} \\ \hline
Exhaustive Search & Global Optimization & \scalebox{0.85}{$\Theta ({2^K}K)$}  \\ \hline
\makecell{Dynamic \\ Programming (DP) } & Global Optimization & \scalebox{0.85}{\makecell{$\Theta (KC) = O(K\overline C ) = $ \\ $ = O({K^2}{c_{\max }})$}}  \\ \hline
\makecell{DP-based \\ Approximation \\ Algorithm (DPAA)} & \scalebox{0.85}{\makecell{$f({{\mathbf{z}}^ * }) \leq f({{\mathbf{\tilde z}}^ * }) \leq $ \\ $\leq f({{\mathbf{z}}^ * }) + \min (\left\lfloor {\epsilon {c_{\max }}} \right\rfloor ,\overline C )$}} & \scalebox{0.85}{\makecell{$O({K^2}\left\lceil {K/\epsilon} \right\rceil ) = $ \\ $ = O({K^3}/\epsilon)$}}  \\ \hline
\end{tabular}
\end{table}

\section{Numerical Simulations and Discussion}
In this section, we examine the performance of the proposed optimization algorithms through numerical simulations. In particular, the following system parameters have been used: $K = 25$ and ${c_k} = \left\lceil {k/5} \right\rceil$, $\forall k \in \mathcal{K}$ ($\overline C  = \sum_{k \in \mathcal{K}} {{c_k}}  = 75$ and ${c_{\max }} = 5$). Moreover, we generate $100$ independent (feasible) optimization problems where the outage probabilities of GSs, ${[{p_k}]_{k \in \mathcal{K}}}$, are uniformly distributed in the interval $(0.25,0.75)$. For the sake of comparison, we consider two \emph{baseline greedy algorithms}, namely, \emph{GD-c} and \emph{GD-p}: first, sort the candidate locations in ascending order of installation cost (respectively, outage probability), and then select the locations $\{1,2,\dots,n\}$ so that $n$ is the smallest integer for which the outage probability threshold is met.

\begin{figure}[!t]
\centering
\includegraphics[width=3.5in]{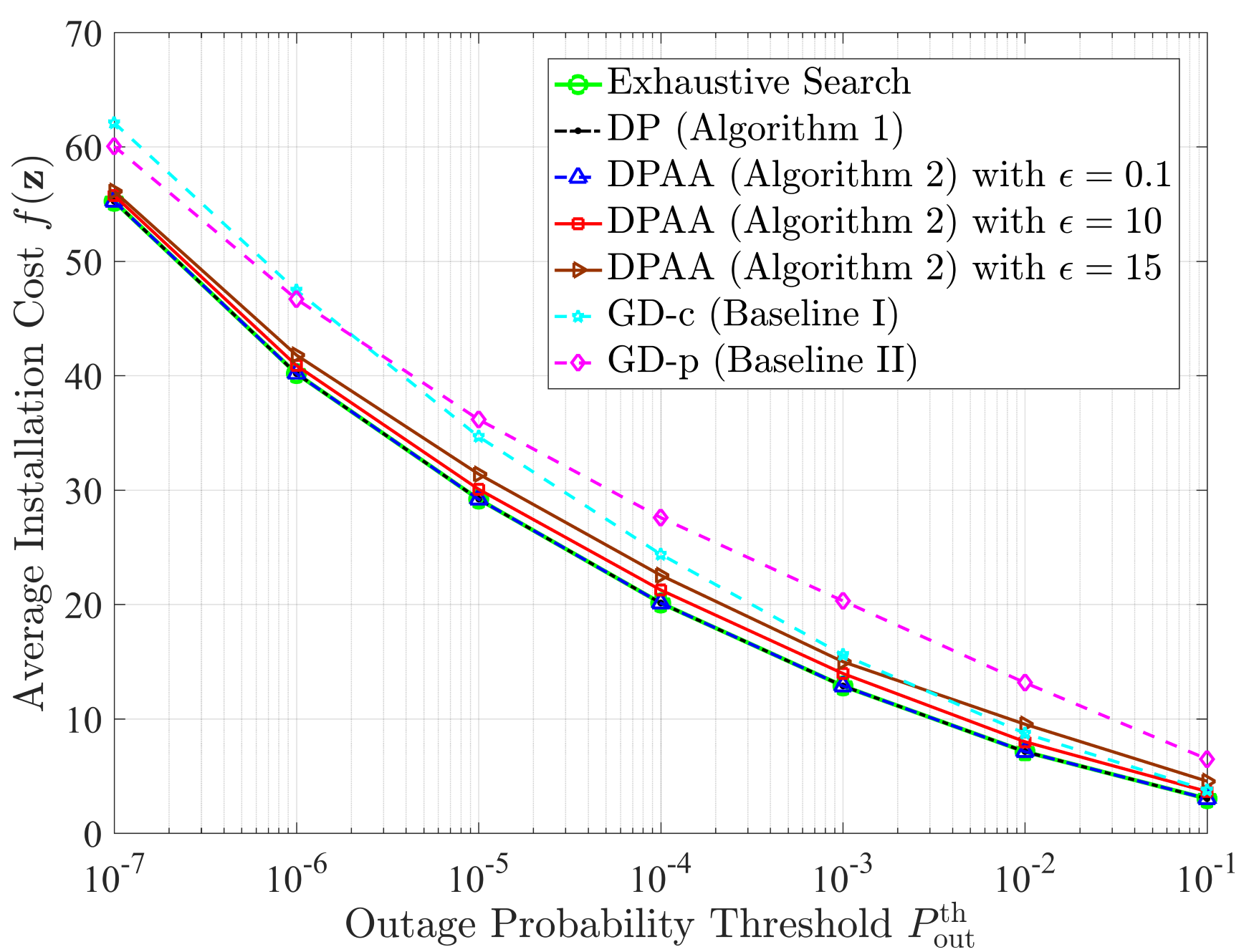} 
\caption{Performance comparison between optimization algorithms.}
\label{Fig1}
\end{figure}

Fig. 1 illustrates the average installation cost, versus the outage probability threshold, achieved by a) the exhaustive search, b) DP (Algorithm 1), c) DPAA (Algorithm 2) for different values of the parameter $\epsilon$, and d) the baseline algorithms. More specifically, DP and DPAA with $\epsilon = 0.1$ have identical performance with the exhaustive search; this is in agreement with the theory presented in the previous sections, since DP is a global optimization algorithm and DPAA is forced to produce an optimal solution when $0 < \epsilon < 1/{c_{\max }} = 0.2$ (see Theorem 3). Furthermore, as expected, DPAA leads to higher installation cost (with lower complexity) by increasing the parameter $\epsilon$. It is interesting to note that, for $\epsilon \in \{10,15\}$, the actual distance of the objective value achieved by DPAA from the global minimum is much less than the absolute-error bound, i.e., $f({{\mathbf{\tilde z}}^ * }) - f({{\mathbf{z}}^ * }) \ll \min (\left\lfloor {\epsilon {c_{\max }}} \right\rfloor ,\overline C )$. Finally, for relatively small outage probability thresholds, the baseline algorithms have lower performance than the proposed algorithms, even for large values of $\epsilon$ (e.g., $\epsilon = \overline{C}/{c_{\max}} = 15$). 

Indicatively, for $P_{{\text{out}}}^{{\text{th}}} = 10^{-4}$ and using a computer with\linebreak Intel Core i7-4790 CPU (3.6 GHz) and 16 GB RAM, the average runtime of the exhaustive search is $2.85$ minutes, whereas that of all the other algorithms shown in Fig. 1 is less than $0.025$ seconds.

\section{Conclusion}
In this letter, we have dealt with the minimization of the installation cost of GSs in RF/optical SatNets satisfying an outage probability constraint. In particular, the examined problem has been theoretically proven to be NP-hard. Moreover, we have presented a global optimization algorithm with pseudo-polynomial complexity as well as a polynomial-time approximation algorithm with provable performance guarantee.

\balance 

\end{document}